**Strong light-matter coupling in lead halide perovskite quantum dot solids**

Clara Bujalance,[1,§] Laura Caliò,[1,§] Dmitry N. Dirin,[2] David O. Tiede,[1] Juan F. Galisteo-López,[1] Johannes Feist,[3] Francisco J. García-Vidal,[3] Maksym V. Kovalenko,[2] Hernán Míguez[1,]*

[1] Multifunctional Optical Materials Group, Institute of Materials Science of Sevilla, Consejo Superior de Investigaciones Científicas – Universidad de Sevilla (CSIC-US), Américo Vespucio 49, 41092, Sevilla, Spain.

[2] Laboratory of Inorganic Chemistry, Department of Chemistry and Applied Biosciences, ETH Zürich, CH-8093 Zürich, Switzerland; EMPA – Swiss Federal Laboratories for Materials Science and Technology, CH-8600 Dübendorf, Switzerland.

[3] Departamento de Física Teórica de la Materia Condensada and Condensed Matter Physics Center (IFIMAC), Universidad Autónoma de Madrid, 28049 Madrid, Spain.

Keywords: Quantum dot solids, Perovskites, Strong exciton-photon Coupling, Polaritons, Optical Microcavities.

**Corresponding Author**

*h.miguez@csic.es.

**Abstract**

Strong coupling between lead halide perovskite materials and optical resonators enables both the polaritonic control of the photophysical properties of these emerging semiconductors and the observation of novel fundamental physical phenomena. However, the difficulty to achieve optical-quality perovskite quantum dot (PQD) films showing well-defined excitonic transitions has prevented the study of strong light-matter coupling in these materials, central to the field of optoelectronics. Herein we demonstrate the formation at room temperature of multiple cavity exciton-polaritons in metallic resonators embedding highly transparent Cesium Lead Bromide quantum dot (CsPbBr$_3$-QD) solids, revealed by a significant reconfiguration of the absorption and emission properties of the system. Our results indicate that the effects of biexciton interaction or large polaron formation, frequently invoked to explain the properties of PQDs, are seemingly absent or compensated by other more conspicuous effects in the CsPbBr$_3$-QD optical cavity. We observe that strong coupling enables a significant reduction of the photoemission linewidth, as well as the ultrafast switching of the optical absorption, controllable by means of the excitation fluence. We find that the interplay of the polariton states with the large dark state reservoir play a decisive role in determining the dynamics of the emission and transient absorption properties of the hybridized light-quantum dot solid system. Our results open the route for the investigation of PQD solids as polaritonic optoelectronic materials.

**Introduction**

Exciton-polaritons arise as a result of the strong coupling between confined photons and bound electron–hole pairs, and are thus characterized by their hybrid light-matter nature.[1-3]

This hybridization takes place within designed optical environments, such as optical cavities, in which the electromagnetic field intensity is magnified for specific photon energies selected to match those of the targeted electronic transitions.[4,5] The exploration of this interaction in the field of lead halide perovskite materials has given rise to polaritonic controlled optical absorption and emission in film-shaped,[6,7] microcrystalline,[8,9] nanosized[10,11] (namely, nanowires,[12] nanoplatelets[13,14] and nanocubes[15]) and low dimensional (such as Ruddlesden-Popper phases[16]) perovskites, which has been put into practice to develop optical switches,[17] lasers,[12,14,18] solar cells,[7] light emitting diodes[9] and photodetectors[19] with enhanced performance. Reciprocally, the integration of these emerging materials in the field of polariton physics has provided the opportunity to observe novel fundamental phenomena.[20-23] However, the observation of strong light-matter coupling in PQD solids has remained elusive, in spite of being some of the most appealing materials for both fundamental analysis and applications in optoelectronics.[24-27] Furthermore, this interaction has been scarcely investigated in QD solids in general, regardless of their composition, with only a few examples employing extremely thin CdSe and CdZnS/ZnS QD films.[28,29] The reason for this is two-fold. First, the difficulty to build QD films of high optical quality (i.e., scattering free), which hinders their integration in an optical cavity, thus avoiding well-defined optical resonances to be achieved. Second, the weak oscillator strengths of excitonic transitions in QD solids at room temperature, which results from the electronic energy disorder originating from the size dispersion of the as-synthesized QDs and, in the case of PQDs, also from the characteristic low exciton binding energy.[30] In this context, recent advances to improve the quality of colloidal PQDs[31] have permitted to attain transparent films made from cubic CsPbBr$_3$ nanocrystals, which have been integrated into a photonic crystal resonator to observe weak light-matter coupling properties, as evidenced by the observation of amplified spontaneous emission, characteristic of the Purcell effect.[32] In another recent achievement, it has been shown that PQD films, whose absorption spectra partially preserve the excitonic features present in the colloidal cubic CsPbBr$_3$ nanocrystals used as building blocks, exhibit the excitonic optical Stark effect under very intense photoexcitation.[33] Both results imply a substantial advancement in the field of perovskite photonics.

In this letter, we demonstrate a robust procedure to achieve strong light-matter coupling between a metallic optical cavity and a PQD solid. Central to this achievement are recent advancements in the preparation of CsPbBr$_3$ QDs with ultra-high monodispersity,[34] which allows to build uniform, large-scale, thick (∼500 nm), transparent PQD films capable of sustaining well-defined excitonic transitions. Following a rational design of the microcavity, hybrid light-matter states are formed, as could be unequivocally confirmed by analysis of the absorption energy dispersion relation. From a practical perspective, the reconfiguration of the electronic and photon states gives rise to a significant reduction of the photoemission linewidth as well as provides the possibility to controllably tune the coupling strength, and thus Rabi splitting, by means of the excitation fluence, with a switching time as fast as one picosecond and a recovery time of the order of a few nanoseconds. Furthermore, analysis of the excitation and relaxation dynamics of the PQD optical cavity indicates that they can be explained without considering the effects of biexciton interaction or large polaron formation, which have been pointed out as partially responsible for the reported ultrafast response of PQDs. Instead, we find that the interplay of the polariton states with the large dark state reservoir, an exclusive feature of hybridized light-matter systems, plays a decisive role in determining the dynamics of the transient absorption and emission properties. Overall, our results lay the ground for future investigations of PQD solids as polaritonic materials and their potential application in optoelectronics.

**Results and discussion.**

CsPbBr$_3$-QDs investigated in this work are capped with long-chain zwitterionic lecithin ligands and synthesized by the recently developed slow-growth room-temperature method, which yields QDs with an unprecedented monodispersity and size-tunability.[34] These CsPbBr$_3$-QDs present a spheroidal rhombicuboctahedral shape with an average diameter of 6.6 nm (Fig. 1a) and show two exceptionally well-resolved excitonic transitions in the absorbance spectrum (Fig. 1d) when diluted with toluene (0.73 mg/mL). Concentrated solutions (88 mg/mL) were spin-coated to prepare solid films after adding polystyrene (PS, 15 wt% with respect to that of CsPbBr$_3$-QDs), which improves the quality and stability of the film (a comparative analysis of the effect of PS is shown in Supplementary Fig. S1). The casted film presents high uniformity and transparency, as shown in the HRTEM image and inset of Fig. 1b, respectively. Essential for the purpose of our work, the absorptance spectra of the film, plotted in Fig. 1e, retains all the well-resolved excitonic transitions present in the colloidal dispersion. From both the colloidal dispersion and the film, intense photoluminescence (PL) emission is observed, with a QY of 77% for the dispersed QDs and 32% for the QD solid (QY = 25% in the case of the CsPbBr$_3$-QDs without the addition of PS, Fig. S1). Then, optical resonators were built by sandwiching the CsPbBr$_3$-QDs transparent solid film between two silver mirrors. The cross-section SEM image displayed in Fig. 1c shows the layered structure of the optical cavity, which consists, sequentially from bottom to top, of a 200 nm thick silver mirror, a 9 nm sputtered silicon nitride protective layer, a 360 nm thick film of CsPbBr$_3$-QDs, and a thermally evaporated 30 nm silver mirror. The cavity length was tuned by varying the thickness of the CsPbBr$_3$-QD film so that the third order resonance of the optical cavity matches the first excitonic transition of the perovskite system (please see Methods section and Fig. S2). A series of absorption spectra measured with unpolarised incident light impinging at 26°, 36° and 46° with respect to the cavity surface normal are depicted in Fig. 1f (purple, red and blue solid lines respectively). The spectra observed indicate a substantial reconfiguration of the absorption properties, with three new absorption peaks whose spectral positions coincide neither with those of the excitonic transitions of the bare CsPbBr$_3$-QDs film (grey dashed line), nor with the underlying cavity modes[35] at the selected angles (vertical dashed lines, more details can be found in the Supplementary Methods, section 1, and in Fig. S2). This can be further confirmed by comparing the absorptance second derivative spectra, plotted in Figs. 1g-1i for all three CsPbBr$_3$-QDs dispersion, film and optical cavity, respectively. Both excitonic and polaritonic transitions are readily recognized as minima in the spectra, and are highlighted by blue and red arrows, respectively. The changes observed in the absorption, along with the significant angular dependence observed, are consistent with the formation of new hybrid states arising from the interplay between cavity photons and CsPbBr$_3$-QDs excitons.

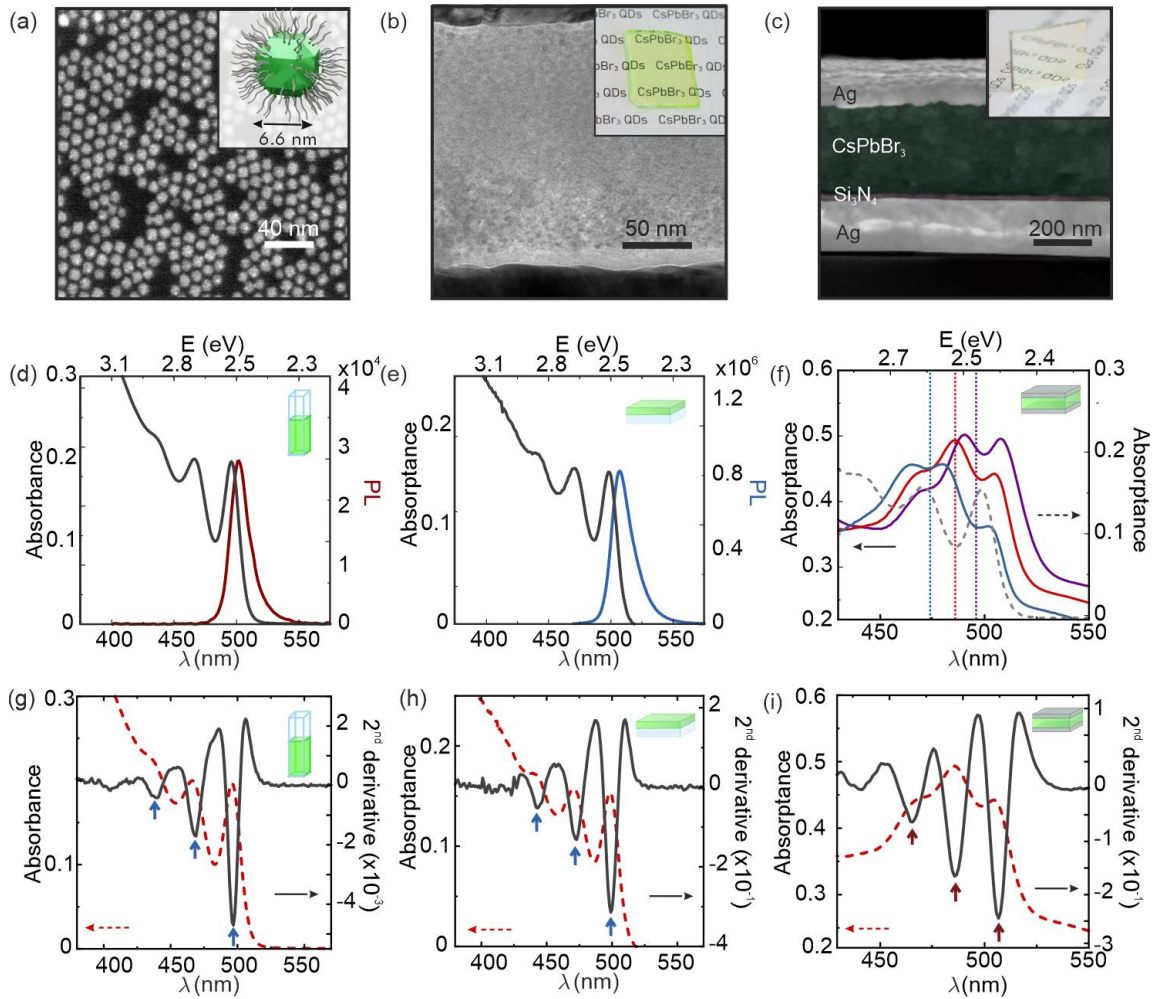

**Figure 1**. **Electron microscopy and linear optical absorption.** (a) TEM image of CsPbBr₃-QDs dispersed on a grid; inset shows the schematic structure of a 6.6 nm lecithin-capped spheroidal CsPbBr₃ nanocrystal. (b) HRTEM image of a spin-casted CsPbBr₃-QD film cross section; inset shows a photograph of such film. (c) SEM image of a cross-section of a CsPbBr₃-QD metallic optical cavity; inset shows a photograph such cavity. (d) Absorbance and photoluminescence (black and red curve, respectively) spectra of a 0.73 mg/mL CsPbBr₃-QD dispersion in toluene. (e) Absorptance and photoluminescence (black and blue curve, respectively) spectra of a CsPbBr₃-QD film. (f) Absorptance spectra of the PQD cavity at angles of incidence 26°, 36° y 46° (purple, red and blue, respectively) attained with unpolarized light. The dotted coloured lines represent the position of the reflectance minima of the third order optical cavity mode at each angle (same colour code); the absorptance of the bare film at 0° (grey dashed line) is plotted for the sake of comparison. (g-i) Second derivative of the absorption spectra (black solid lines) compared with absorptance spectra (red dashed lines) of CsPbBr₃-QDs (g) dispersion, (h) thin film and (i) cavity. Blue and red arrows indicate the minima in the second derivatives that correspond to excitonic and polaritonic transitions, respectively.

Energy dispersion absorption maps are attained by plotting the absorptance versus the photon energy (y-axis) and the parallel component of the wavevector $\mathbf{k}_\parallel$ (x-axis, $\mathbf{k}=\mathbf{k}_\parallel+\mathbf{k}_\perp$) for the two polarizations of the incident beam studied, namely transversal electric (TE, Fig. 2a) and transversal magnetic (TM, Fig. 2b). For the sake of comparison, simulated maps (left panels in each figure) calculated using the transfer matrix method (please see Methods section) are plotted alongside the experimental ones (right panels), showing a fair agreement. The optical constants of the PQD solid used for the calculations were determined experimentally from the transparent films and are shown in Fig. S3. The dispersion of the third order cavity mode is drawn as a green dotted line, while horizontal white dotted lines indicate the positions of the first (1s-1s, $\hbar\omega_s$), second (1p-1p, $\hbar\omega_p$) and third (1d-1d, $\hbar\omega_d$) excitonic transitions of the CsPbBr$_3$-QD film.[34] Analysis of the results shown in Fig. 2 confirms that the strong coupling between exciton and photon modes leads to the formation of hybrid light-matter states in the ensemble. These are featured by three clearly distinguishable polaritonic branches (lower, middle and upper polaritons: LP, MP and UP) and the opening of energy gaps resulting from the anti-crossing of such branches. Energy gaps are estimated at the crossing point between the cavity dispersion curve and the exciton transition position, and, consistently, present a value of $\hbar\Omega_{R1}$ = 87.1 meV between LP and MP and of $\hbar\Omega_{R2}$ = 86.2 meV between the MP and UP, where $\hbar$ is the reduced Planck constant and $\Omega_R$ is the Rabi frequency, which stands for the rate at which energy is exchanged between the photonic mode and the excitonic transitions. The magnitude of these gaps, also known as Rabi splittings, depend on the number of QDs effectively contributing to the coupling ($N$), the oscillator strength of the targeted transition ($f$) and the effective cavity volume occupied by the optical resonance involved in such coupling ($V_C$) as:

$$\Omega_R \sim \sqrt{Nf/V_C} \qquad (1)$$

In fact, cryogenic experiments demonstrate that $\hbar\Omega_{R1}$ is gradually enlarged up to almost 30% due to the increase of $f$ with decreasing temperature, caused by the reduction of phonon scattering, reaching $\hbar\Omega_{R1}$ = 112 meV at 80K, as shown in Supplementary Fig. S4. Further insight into the PQD solid-cavity coupling may be obtained by solving the Tavis-Cummings Hamiltonian, $\hat{H}_{TC}$, which describes a system of $N$ non-interacting particles coupled to a single cavity light mode in the low excitation limit.[35-36] Within this framework, we can obtain the theoretical polariton dispersions, plotted as white solid lines in Figure 2, which show a very good agreement with those experimentally estimated, and the Hopfield coefficients, $C_\gamma$, $C_s$, and $C_p$, whose squared values give us the degree of contribution to the coupling of the cavity mode and of each excitonic transition, respectively. These coefficients are plotted versus the angle of incidence of incoming light with respect to the cavity normal in Figs. 2c-2e. From these, it can be readily seen that LP and UP are mainly participated by the cavity mode and either the first (1s-1s) or the second (1p-1p) excitonic transition, respectively, while the MP shows a much even contribution from all states. In Fig. S5, we plot the calculated absorptance, as well as the spatial and spectral profiles of both the intensity of the electric field, $|\mathbf{E}(\mathbf{r})|^2$, and the corresponding absorbed luminous power, P$_A$, along a cross-section of a CsPbBr$_3$-QDs filled optical cavity for those angles for which the respective contribution of the optical mode and the excitonic transitions to the coupling is approximately 50% (highlighted as vertical dashed lines in Figs. 2c-e). Full details on the solution of the $\hat{H}_{TC}$ are provided in the Supplementary Methods Section 1. It should be remarked that the vast majority of solutions, whose energies lie in between those of the LP, MP and LP, cannot be accessed from the ground state for symmetry reasons and are referred to as dark states. Interestingly, although this large reservoir of dark states is not observable in a linear absorption experiment, it significantly contributes to determine the dynamics of both the ultrafast transient absorption and PL decay properties of the PQD cavities, as it will be shown further below.

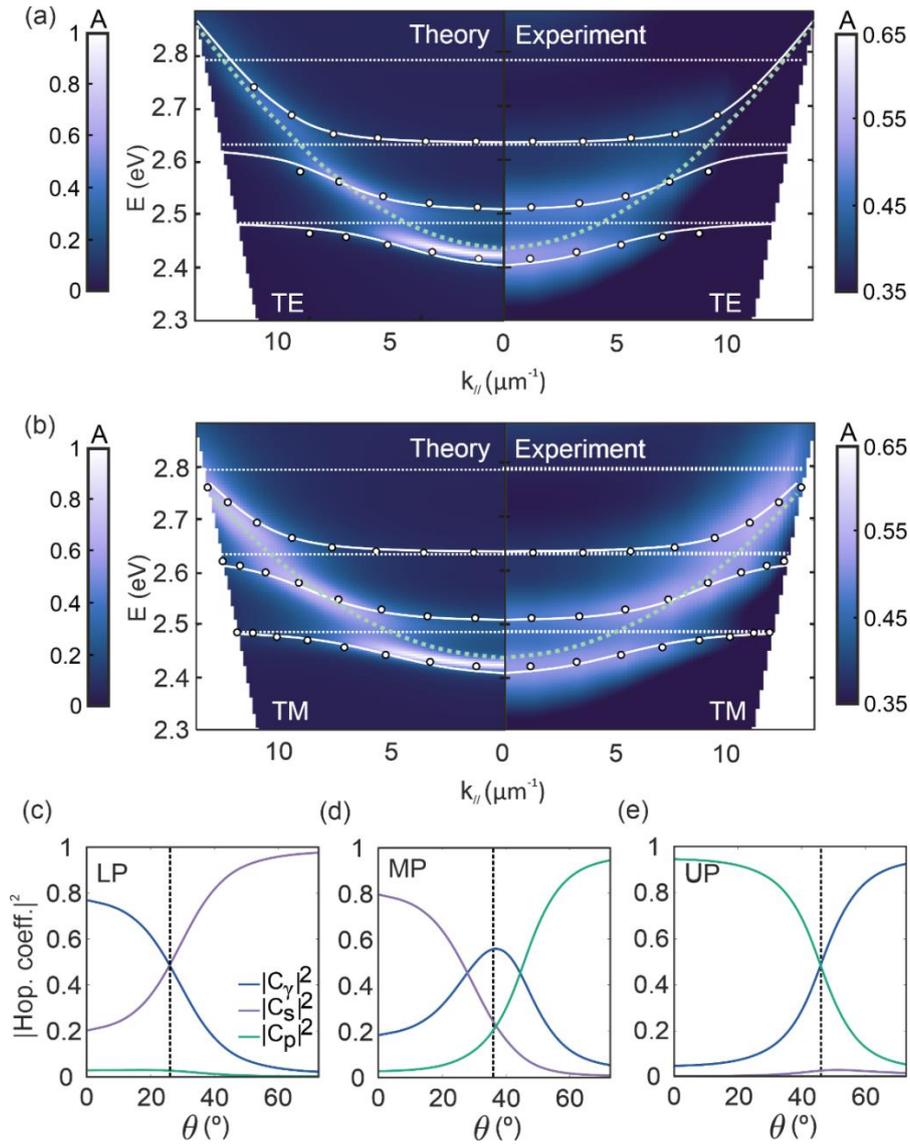

**Figure 2**. **Exciton-polariton energy dispersion relation for a CsPbBr₃-QD cavity**. Theoretical (left panel) and experimental (right panel) absorptance energy dispersion maps for (a) TE and (b) TM polarisations of the incident beam. White dots indicate the spectral position of the experimental absorption maxima, horizontal dotted white lines indicate the positions of the first three excitonic transitions, green dotted lines represent the underlying cavity mode dispersion and solid white lines are the LP, MP and UP dispersions attained by solving the Tavis-Cummings Hamiltonian. (c-e) Angular dependence of the squared Hopfield coefficients ($|C_\gamma|^2$, $|C_s|^2$ and $|C_p|^2$, plotted as blue, purple and green lines, respectively) attained from the three coupled oscillator model for the (c) LP, (d) MP and (e) UP; vertical dashed lines indicate the angles at which the photon and exciton contribution to the coupling have a similar weight for each polariton.

Polariton excitation and decay dynamics in CsPbBr₃-QD optical microcavities were studied by ultrafast transient absorption spectroscopy (TAS). Results are presented as intensity maps in line, right y-axis) as extracted from the fittings shown in (e) and (f). The green and grey shaded regions in Fig. 3f highlight the duration of the excitation pulse, $\Delta t$=190 fs, and the estimated time for carrier thermalization, $\Delta t \approx$ 350 fs, respectively.[37]

Fig. 3a, in which the full series of ΔA spectra (where ΔA is the result of subtracting the absorption of the non-excited cavity from the photoexcited one) are plotted as a function of probe photon wavelength and pump-probe delay, $\Delta t$, following the non-resonant excitation (i.e., not matching the polaritonic transitions in the cavity) at time 0 by a 190 fs, $\lambda$=420 nm and ≈120 μJ/cm² laser pulse. Under these conditions, we obtain $N_{e-h}$≈0.6 excitations per QD, preventing the damage of the samples. Selected ΔA spectra attained at different delay times, namely at $\Delta t$=0.4 ps (green line) and $\Delta t$=10 ps (purple line), are explicitly shown in Fig. 3b, while the early stage dynamics of the main signals identified in Fig. 3a are plotted in Figs. 3c. For the sake of comparison, a similar analysis, which is provided in the Supporting Information (Fig. S6 and Supplementary Methods, section 5) was performed for the CsPbBr₃-QD colloidal dispersion and the bare film. The geometry of the experiment employed for each type of sample is described in the Methods section.

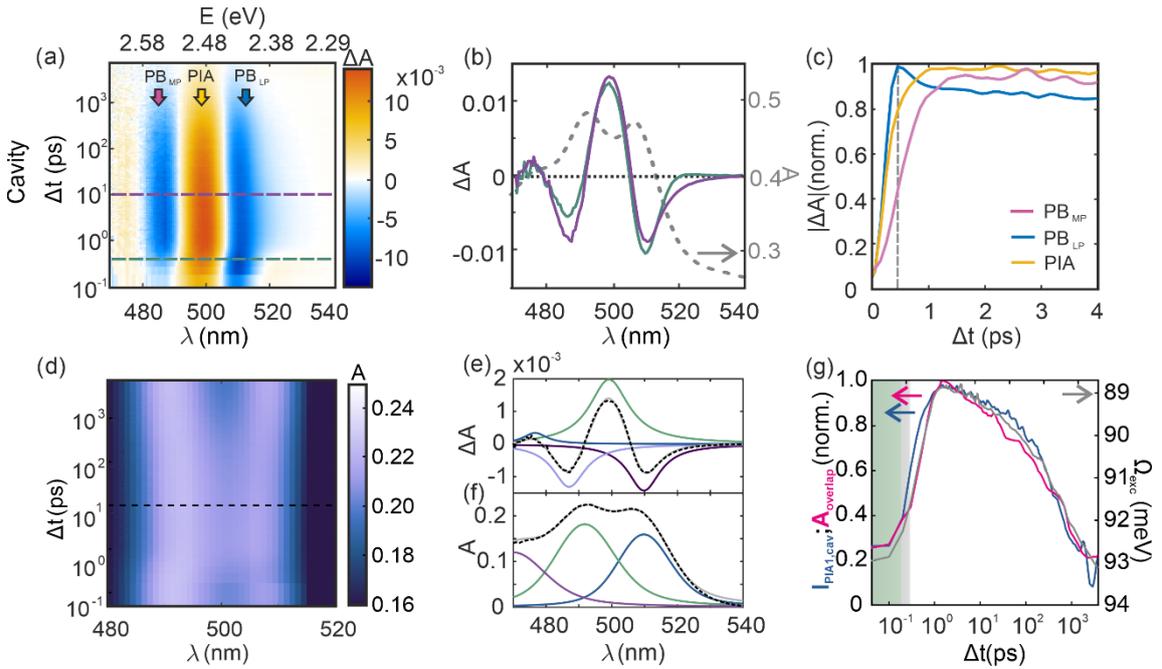

**Figure 3**. **Ultrafast transient absorption spectroscopy of CsPbBr₃-QD film and optical cavity and ultrafast switching of the Rabi splitting.** (a) Map of ΔA versus $\lambda$ and $\Delta t$ for the CsPbBr₃-QD optical cavity, in which the main signals are labelled. (b) Some illustrative selected ΔA spectra, attained at $\Delta t$=0.4 ps (green line) and $\Delta t$=10 ps (purple line); both $\Delta t$ are highlighted by dashed horizontal lines in (a). The corresponding linear absorptance spectra are also plotted (grey dashed lines). (c) Early-time evolution of the maximum intensity of the main signals extracted from the analysis of |ΔA|. (d) Map of the reconstructed linear absorptance versus $\Delta t$ and $\lambda$. (e) Illustrative example of the deconvolution of a selected ΔA spectrum ($\Delta t$=10 ps, dashed horizontal line in panel (a) into its main component signals: PB$_{LP}$, PB$_{MP}$ and PIA (dark violet, light violet and green solid lines, respectively); grey solid curve shows the sum of all three contributions. (f) Illustrative example of the deconvolution of a selected TAS signal modulated absorptance spectrum into a LP, MP and UP absorption peaks (blue, green and violet solid lines, respectively). Experimental curves and fittings are also plotted (black dashed and grey solid lines, respectively). (g) Comparison between the evolution of the maximum intensity of I$_{PIA}$ and A$_{overlap}$ (blue and pink solid lines, respectively, left y-axis) and that of $\Omega_{exc}$ (grey solid

Two prominent photobleaching signals are detected in Fig. 3a at the lower (PB$_{LP}$) and middle (PB$_{MP}$) polariton spectral positions, which show a very different rise time ($\tau_{PB,LP}$<0.5 ps, $\tau_{PB,MP}$<2 ps) as it can be seen in Fig. 3c (blue and pink lines). In this regard, while the bleaching of the LP may be attributed to filling of the lowest energy states due to hot-electron relaxation, the subsequent long-lived bleaching of the MP cannot be understood without considering the interplay with the large reservoir of dark states. Even though not directly accessible from the ground state, dark states can be filled both from the lower polariton state and from higher energy levels as hot electrons cool down.[38] This picture is further supported by the partial recovery of the absorption evidenced by the peak observed in the PB$_{LP}$ signal at $\Delta t$=0.45 ps (signposted by a vertical grey dashed line in Fig. 3c), which points at a transfer of carriers from the lower polariton state to the dark state reservoir once a certain occupation level is attained. Thus, gradual filling of the dark states could eventually lead to transfer of carriers to the middle polariton state, giving rise to its bleaching. These carrier exchanges are favoured by the significant overlap between dark and polaritonic states,[39,40] which can be inferred from the moderate separation between lower and middle polaritons we observe.

As for the intense PIA signal observed between PB$_{LP}$ and PB$_{MP}$, the detailed analysis of the TAS signal indicates that it is the result of the photoinduced reduction of the polariton absorption splitting. In fact, non-resonant photo-pumping could give rise to the inhibition of the Rabi splitting as a consequence of the screening of the exciton transition dipole moment caused by the presence of photogenerated electron-hole pairs, as expressed in the formula:[41]

$$f = \frac{f_0}{(1+\frac{N_{e-h}}{N_S})} \qquad (2)$$

In which $f_0$ is the oscillator strength in the absence of pumping, and $N_{e-h}$ and $N_s$ are the pump generated and saturation electron-hole pair densities, respectively. Considering the relation between $f$ and the $\Omega_R$ given by equation (1), the absorption splitting of a polaritonic system under intense photoexcitation ($\Omega_{exc}$) is reduced, from its linear absorption value $\Omega_0$, to:

$$\Omega_{exc} = \frac{\Omega_0}{\sqrt{1+\frac{N_{e-h}}{N_S}}} \qquad (3)$$

The narrowing of $\Omega_{exc}$ becomes explicit when we plot the result of modulating the linear absorptance with the $\Delta A$ data attained from TAS at all $\Delta t$, as shown in Fig. 3d. In this representation, the intense PIA signal observed may be understood as the result of the smaller energy separation between LP and MP absorption peaks, which gives rise to newly available states in the spectral region in which there was a gap initially, thus opening the possibility to observe new transitions from the ground state. To support this hypothesis, from the fitting of the TAS signal at all $\Delta t$ (an example is shown in Fig. 3e for $\Delta t$=10 ps) and that of the reconstructed absorptance spectra (as exemplified in Fig. 3f, also at $\Delta t$=10 ps), we extract the time evolution of both the PIA signal and $\Omega_{exc}$, estimated as the spectral difference between the maxima of the two peaks corresponding to the transitions from the ground state to the lower and middle polaritons ($A_{LP}$ and $A_{MP}$). The comparison of the PIA signal intensity, $I_{PIA}$, and $\Omega_{exc}$ versus $\Delta t$ is shown in Fig. 3g (grey and blue lines, respectively). The time dependence of the overlapping area of $A_{LP}$ and $A_{MP}$, $A_{overlap}$, is also plotted (pink solid line, Fig. 3g). This plot reveals a strong correlation between the splitting reduction and the occurrence of the observed PIA, hence further supporting their causal link. Also, from this analysis and equation (3), we can estimate a saturation $e^-$-$h^+$ density per QD of $N_S \approx 3.5$, considering that $\Omega_0 \approx 93$ meV, and that $\Omega_{exc} \approx 89$ meV for an $e^-$-$h^+$ density $N_{e-h} \approx 0.6$.

It should be remarked that a similarly pronounced PIA, spectrally located between the 1s-1s and 1p-1p excitonic transitions, has been reported for CsPbBr$_3$-QDs dispersions,[42,43] and it is also observed in our bare films (Fig. S6). Its origin has been attributed to either the relaxation of selection rules due to the formation of large polarons,[44,45] which might enable parity forbidden exciton transitions,[42] or to higher energy exciton-biexciton transitions.[43] Interestingly, none of these phenomena seem to be playing a significant role in the ultrafast response of the photon dressed excitons formed in the cavity. From a more applied perspective, these results demonstrate the possibility to attain ultrafast switching of the exciton-polariton absorption, an effect that had only been shown for molecular polaritons coupled to surface plasmon polaritons,[46] for which the duration of the Rabi splitting reduction was three orders of magnitude shorter than in our case, in which the linear absorptance is fully recovered (thus the PIA disappears) at $\Delta t \approx 10$ ns, when all the photoexcited $e^-$-$h^+$ pairs decay back to the ground state.

The photoemission of the strongly coupled PQD solid-cavity system also undergoes a significant spectral and directional reconfiguration. In Fig. 4, the PL spectra measured from either a CsPbBr$_3$-QD film on quartz (Fig. 4a) or from a strongly coupled CsPbBr$_3$-QD cavity (Fig. 4b) are shown. Measurements were taken using a back focal plane spectroscopy set up, which allows us to collect the emitted radiation in a wide angular range (-30°<$\theta$<30°) by scanning the Fourier plane with an optical fibre coupled to a CCD detector, i.e., without tilting the sample and hence assuring that all spectra are collected from the same spot and under the same photoexcitation intensity from a continuum diode laser emitting at $\lambda$=450 nm. Analysis of these measurements shows that the CsPbBr$_3$-QD cavity is determined by the spectral dispersion and the linewidth of the lower polariton transition[47] as it can be explicitly seen in the comparison between the polariton absorption and emission made in Fig. 4c. In this case, absorptance was estimated also using back focal plane spectroscopy. The horizontal dashed line indicates the spectral position of the first uncoupled excitonic transition, while the white solid line curves shows the angular dispersion of the LP and MP. A very significant 35% reduction of the spectral photoemission linewidth, from 96 meV in the film to 62 meV in the cavity, is observed. Emissions occurring from middle and upper polaritons were not detected, since higher energy polariton states have a very short lifetime due to non-radiative decay to the large number of incoherent states present in the dark state reservoir.[48]

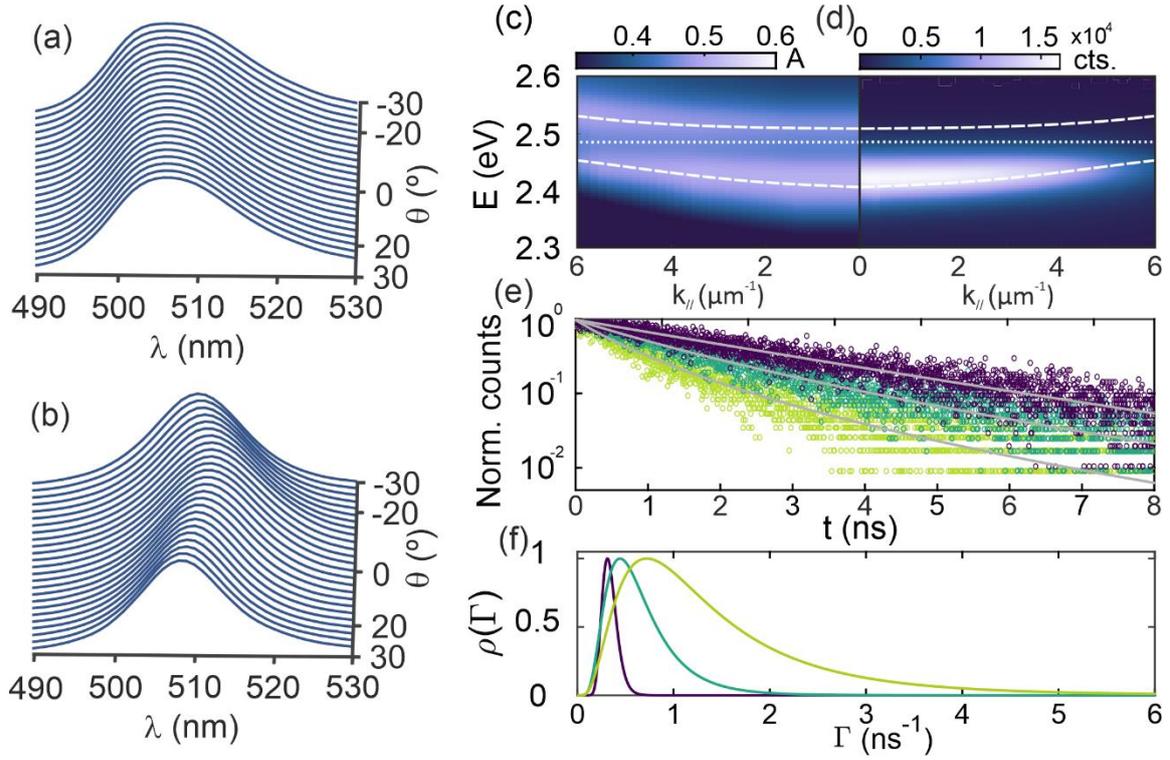

**Figure 4. Static and dynamic photoemission properties of CsPbBr₃-QD dispersion, film and optical cavity**. Normalized PL spectra at different collection angles (from -30° to 30°) of the (a) bare film and (b) the optical cavity. Comparison of the optical cavity (c) absorptance and (d) PL dispersions along $k_{//}$. The white dotted line indicates the spectral position of the original excitonic transition, and the white dashed lines are the theoretical results attained by solving the Tavis-Cummings Hamiltonian. (e) PL decay curves attained for the CsPbBr₃-QD colloidal dispersion, film and optical cavity (violet, dark green and light green open circles, respectively) together with their corresponding fittings according to a log-normal distribution of decay rates (grey solid lines). For the sake of comparison, also in this case, data were obtained for a similar number of photoexcitations per CsPbBr₃-QD ($N_{exc} \approx 0.016$). (f) Decay rate distributions, $\rho(\Gamma)$, attained for the best fittings of the PL decay curves for the CsPbBr₃-QDs dispersion, film and optical cavity (violet, dark green and light green solid lines, respectively).

The dynamic properties of the excited states were studied by probing the time evolution of the emission with a time-correlated single photon counting (TCSPC) set-up. The same ultrafast pulsed laser employed for TAS measurements emitting at $\lambda$=420 nm was employed to photoexcite the samples, whose emission was collected with an avalanche photodiode. In Fig. 4e we show PL decay curves attained for all three systems under study, together with their corresponding fits assuming a log-normal distribution of decay rates (grey solid lines). For the sake of comparison, we have plotted a set of data obtained for similar number of photoexcitations per CsPbBr₃-QD ($N_{exc} \approx 0.016$, be it in suspension, in the film or as part of the optical cavity). The full collection of PL decay curves measured for different fluences for each sample can be consulted in Supplementary Fig. S5. Analysis of the decay rate distributions, $\rho(\Gamma)$, which are drawn in Fig. 4f, reveals that the CsPbBr₃-QD film presents a much wider range of decay paths, with respect to the colloid, as the much narrower $\rho(\Gamma)$ of the latter indicates. Accordingly, the most likely lifetime ($\tau$, corresponding to the $\Gamma$ for which $\rho(\Gamma)$ is maximum,

$\Gamma_{max}$) is also significantly higher in the CsPbBr$_3$-QDs dispersion than in the film, which implies that not only the variety, but also the density of non-radiative states increases in the film. The presence of a higher density of non-radiative pathways in the film is likely the consequence of the formation of multiple interfaces due to the CsPbBr$_3$-QDs packing. In turn, the polaritonic system presents an even broader $\rho(\Gamma)$ and shorter characteristic $\tau$ (or higher $\Gamma_{max}$) than the film, which agrees with the presence of a large reservoir of non-radiative dark states to which photoexcited carriers can be transferred from the LP level. These results are therefore also in line with those attained from the analysis of the TAS signal, which also supports that relaxation dynamics strongly depend on the interplay with the dark states.

In conclusion, we have demonstrated strong light-matter coupling between a CsPbBr$_3$-QD solid film and an optical cavity. Central to this achievement is the preparation of scattering-free films, made of highly monodisperse CsPbBr$_3$ nanocrystals, displaying up to three well-resolved excitonic transitions in the absorption spectra. This allows us to fabricate metallic optical resonators with enough quality as to give rise to the formation of three polaritonic branches (namely upper, middle and lower), as a result of the superposition of two excitonic and one optical cavity excitations, as could be unequivocally confirmed by analysis of the absorption energy dispersion relation. Comparative analysis of both the ultrafast transient absorption and the photoluminescence decay of all three CsPbBr$_3$-QDs systems under analysis (i.e., dispersion, film and optical cavity), reveals a very different dynamics in the case of the photon-dressed excitons in the resonator with respect to bare ones in the CsPbBr$_3$ nanocrystals. Our results indicate that the effects of biexciton interaction or large polaron formation, frequently invoked to explain the transient absorption properties of PQDs, are seemingly absent or compensated by other more conspicuous effects in the CsPbBr$_3$-QDs optical cavity. Instead, we find that the interplay of the polariton states with the large dark state reservoir plays a decisive role in determining the dynamics of the transient absorption and emission properties of the hybridized light-matter system. From a practical perspective, the reconfiguration of the electronic and photon states gives rise to a significant reduction of the photoemission linewidth as well as provides the possibility to controllably tune the coupling strength, and thus Rabi splitting, by means of the excitation fluence, with a switching time as fast as one picosecond and a recovery time of the order of a few nanoseconds. Our results should serve as the basis for future investigations of PQD solids as polaritonic materials and of their potential application in optoelectronics.

**Experimental Methods**

*Colloidal CsPbBr$_3$ quantum dot synthesis.*

Lead bromide (PbBr$_2$, 99.999%), Cesium carbonate (Cs$_2$CO$_3$, 99.9%), hexane (≥99%), diisooctylphosphinic acid (DOPA, 90%), and acetone (ACE≥99.5%) were purchased from Sigma Aldrich. n-Octane (min. 99%) and lecithin (>97% from soy) were purchased from Carl Roth. Trioctylphosphine oxide (TOPO, >90%) was purchased from Strem Chemicals. All chemicals were used as received.

PbBr$_2$-TOPO stock solution (0.04 M) was prepared by mixing 4 mmol of PbBr$_2$ with 20 mmol of TOPO in 20 mL of n-octane at 120 $^o$C. The resulting solution was later diluted with 80 mL of hexane. Similarly, the Cs-DOPA solution (0.02 M) was prepared by mixing 100 mg of $Cs_2CO_3$ with 1 mL of DOPA in 2 mL of n-octane at 120$^o$C, and subsequently diluted in 27 mL of hexane. A 0.13 M lecithin stock solution was prepared by dissolving 1.0 g of lecithin in 20 mL of hexane. All stock solutions were filtered through 0.2 μL PTFE before the use.

For the synthesis of 6.6 nm-large $CsPbBr_3$-QDs, 30 ml of $PbBr_2$-TOPO stock solutions were diluted with 180 ml of hexane, followed by the injection of 15 ml of Cs-DOPA stock solution under vigorous stirring. After 4 minutes of growth, 15 ml of lecithin stock solution were added. After a minute, the obtained solution was concentrated to 15 ml on a rotary evaporator and a 3-times volume excess of acetone acting as antisolvent was added. QDs were isolated by centrifuging at 20133 g for 1 minute and re-dispersed in 24 ml toluene.

An additional washing with toluene/ethanol pair of solvent/antisolvent was performed in order to remove the excess of lecithin. QDs were precipitated from the crude solution by adding 24 ml of ethanol and centrifuging the mixture at 20133 g for 1 minute. The product was redissolved in 12 ml toluene and washing was repeated with 12 ml ethanol, followed by redissolution in 6 ml toluene. On the third washing cycle, QDs were precipitated by 6 ml and, after centrifugation, redissolved in 2 ml toluene. The obtained solution contains 88 mg/ml of QDs.

*CsPbBr$_3$ QD Film and cavity preparation.*

Chemical reagents and solvents (i.e., polystyrene (PS) and toluene) were purchased from Sigma–Aldrich (highest grade available) and were used without further purification. A 2 wt% solution of PS in toluene was stirred at room temperature until complete dissolution of the polymer; then, the appropriate amount was added to the 88 mg/mL QDs dispersion in toluene in order to obtain a 15wt% of PS with respect to the QDs. The resulting QDs/PS mixture in toluene was then spin-coated on the substrate by varying the spin-coating speed in order to adequately tune the cavity thickness. In order to prepare the cavity, substrates with 200 nm of silver and 9 nm of sputtered silicon nitrate were purchased by Fraunhofer, and cleaned by ultrasonic bath (with 2% Hellmanex, acetone and 2-propanol) followed by 10 min treatment with oxygen plasma. Then, on top of the $CsPbBr_3$-QDs spin-coated layer, a silver mirror of 30 nm was thermally evaporated in vacuum at $10^{-6}$ mbar (Univex 250, Leybold vacuum), at 1 Å/s until reaching the desired thickness, by monitoring it with a quartz balance coupled to the system.

*Linear absorption characterization and optical constants determination.*

Polarization and angular dependent measurements were conducted with double goniometer configuration (Universal Measurement Accessory, UMA), which allows us to rotate independently sample and detector and hence select arbitrary incident and collection directions, both attached to a UV-Vis-NIR spectrophotometer (Cary 5000, Agilent). Absorptance (A) was attained as A = 1 − R − T, where R and T stand for reflectance and transmittance, respectively. The optical constants for the QD solid film were estimated by the method developed by Forouhi and Bloomer,[49] based on the fitting of the experimental T and R measured at different angles of incidence and polarizations with the UMA. With this approach, we could estimate the spectral dependence of both real and imaginary components of the refractive index for the bare PQD solid film, which were then used to design the geometry of the optical cavity (see Fig. S3 of Supplementary Information).

*Static and dynamic photoluminescence analysis*

Static photoluminescence was measured by back focal plane spectroscopy, using a Leica DMI300M microscope with a 100x, 0.75NA objective. The image from the sample's back focal plane was directed to an optical fiber mounted on a motorised stage scanning the horizontal axis, each position of the fiber corresponding to a given angle of detection. The source of excitation is a 450 nm continuous laser and signal is detected by a UV-VIS spectrophotometer (Ocean Optics. USB200).

Lifetime measurements were carried out with a TCSPC set-up using an Ytterbium-doped potassium gadolinium tungstate (Yb:KGW, $\lambda$=1040 nm) femtosecond pulsed laser (PHAROS PH1 from Light Conversion) operating at 1KHz (pulse duration, 190 fs) sending 420 nm pulses and collecting the time resolved photoluminescence with a single photon avalanche photodiode from MPD. The signal was processed employing commercial software provided by Ultrafast Systems.

*Ultrafast spectroscopy analysis.*

Ultrafast transient absorption spectroscopy (TAS) measurements were performed using a pump-and-probe setup. The signal generated by a Yb:KGW pulsed laser is split into two beams. One is directed to an optical parametric amplifier (Orpheus, Light Conversion) where 420 nm pump pulses with a narrow spectral width (4 nm) are generated. The other beam passes through a delay line and, subsequently, a sapphire crystal, to generate broadband probe pulses. Signal was collected with a CMOS detector attached to a spectrometer (HELIOS, Ultrafast Systems) in a transmission configuration for transparent samples (i.e., nanocrystal dispersions and thin films on quartz) and in a reflection configuration (with an angle of incidence and collection of the probe beam of 26˚) for opaque samples (metallic optical cavity). Collected signal is the result of TAS data are initially processed with Surface Xplorer software (background and chirp corrections) and further analysis (Voigt fit of transient absorption spectra) is performed with a Matlab code (See Supplementary Methods 4). A constant scattering background was removed from the measurements before performing the fittings. The HELIOS Fire software produces a 3D wavelength-delay time-$\Delta$A matrix, where $\Delta$A represents a differential absorbance, as it is calculated as the logarithm of the ratio of the light non-absorbed by the ground state and the photoexcited sample, namely $\Delta A = -\log\frac{I_0}{I_{exc}}$. The density of e⁻-h⁺ per unit volume generated in the samples with the pump pulses are calculated assuming that the excitation intensity over the volume of the sample is the average between the maximum intensity and the intensity attenuated after the absorption over one length of the sample (See Supplementary Methods 3).[50]

*Optical Simulations*

Optical reflectance, transmittance, and absorptance, as well as electric field intensity and absorption profiles, are calculated with a Matlab code using a transfer matrix method based on the Abeles formalism.[51] The thickness of each layer is determined experimentally using a profilometer and the refractive indices are either taken from bibliography (Ag, $Si_3N_4$) or modelled through the fitting of the experimental R and T spectra using a Forouhi-Bloomer model of the dielectric constant ($CsPbbr_3$-QDs layer).[49] The spatial distribution of the EM field intensity along the direction of propagation at $0^o$ ($z$), calculated in small intervals, is used to obtain the absorption per unit volume. This value, integrated in $z$ between the limits imposed by the interfaces between layers, yields the theoretical absorptance of each layer in the ensemble.

## Acknowledgements

H.M. is thankful for the financial support received by the Spanish Ministry of Science and Innovation-Agencia Estatal de Investigación (MICIN-AEI) under grants PID2020-116593RB-I00 and TED2021-129679B-C22, funded by MCIN/AEI/ 10.13039/501100011033 and by Unión Europea NextGenerationEU/PRTR, by the Junta de Andalucía under grant P18-RT-2291 (FEDER/UE) and the Innovative Training Network Persephone ITN, funded by the European Union's Horizon 2020 research and innovation programme under the Marie Skłodowska-Curie grant agreement No 956270. M. V. K. and D.D. acknowledge financial support from the Air Force Office of Scientific Research and the Office of Naval Research (award number FA8655-21-1-7013). J.F. and F.J.G.-V. acknowledge support by the Spanish MICIN–AEI under grants PID2021-125894NB-I00 and CEX2018-000805-M (through the Mara de Maeztu program for Units of Excellence in Research and Development). L.C. thanks Junta de Andalucía and the European Regional Development Funds program (EU-FEDER) for financial support under a talent attraction program (DOC_00220). We are grateful to Maryna I. Bodnarchuk for the interesting discussions.

## Data availability

The data underlying this study are openly available in the Digital CSIC repository. The codes used for generating absorptance spectra and the spatial distribution of the optical field intensity and the absorption in an optical cavity are provided at https://github.com/Multifunctional-Optical-Materials-Group.

## Author contribution

M.V.K. and H.M. conceived the original idea of preparing polaritonic quantum dot solids based on transparent films and supervised the work. D.D. synthesized the $CsPbBr_3$ quantum dots. C.B. and L.C. prepared the optical quality films and the resonators, and performed all the linear optical characterization. D.O.T. and J.F.G. performed the ultrafast TAS measurements. J.F. and F.J.G-V. provided theoretical insight on the polaritonic effects observed. C.B. and L.C. prepared the figures. L.C. and H.M. wrote the first manuscript draft. All authors contributed to the discussion, analysis and writing of the final version of the paper.

# Strong light-matter coupling in lead halide perovskite quantum dot solids

*Clara Bujalance,[1,§] Laura Caliò,[1,§] Dmitry N. Dirin,[2] David O. Tiede,[1] Juan F. Galisteo-López,[1] Johannes Feist,[3] Francisco J. García-Vidal,[3] Maksym Kovalenko,[2] Hernán Míguez[1,]\**

*1 Multifunctional Optical Materials Group, Institute of Materials Science of Sevilla, Consejo Superior de Investigaciones Científicas – Universidad de Sevilla (CSIC-US), Américo Vespucio 49, 41092, Sevilla, Spain.*

*2 Laboratory of Inorganic Chemistry, Department of Chemistry and Applied Biosciences, ETH Zürich, CH-8093 Zürich, Switzerland; Empa – Swiss Federal Laboratories for Materials Science and Technology, CH-8600 Dübendorf, Switzerland.*

*3 Departamento de Física Teórica de la Materia Condensada and Condensed Matter Physics Center (IFIMAC), Universidad Autónoma de Madrid, 28049 Madrid, Spain.*

# Table of Contents

Supplementary Figures

Supplementary Methods

**Supplementary Figures**

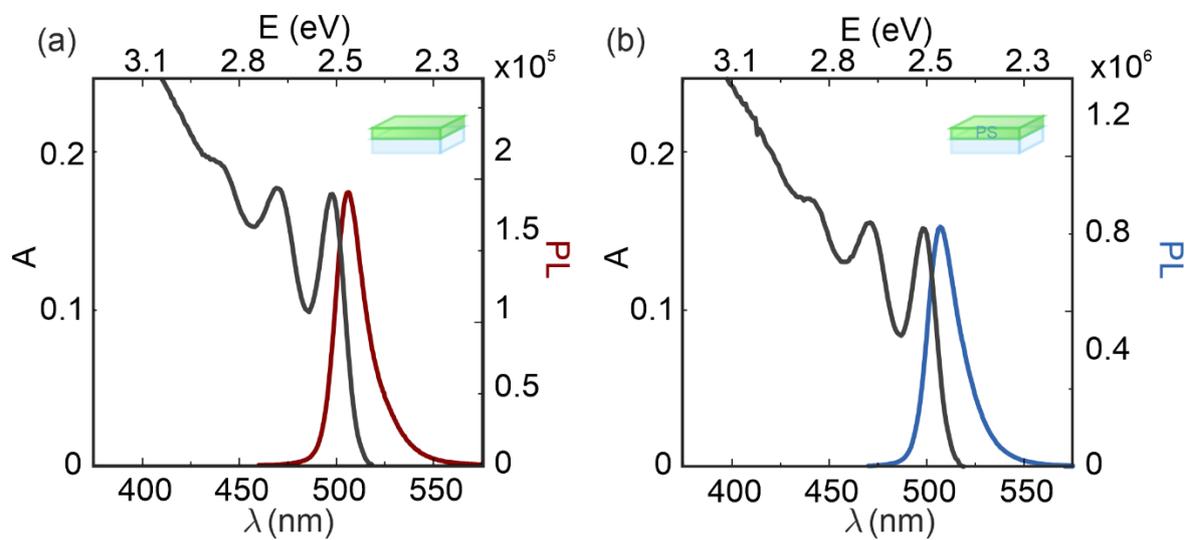

**Figure S1.** Absorptance(black) and photoluminescence (red/blue) spectra of the CsPbBr₃ film without (a) and with (b) polystyrene.

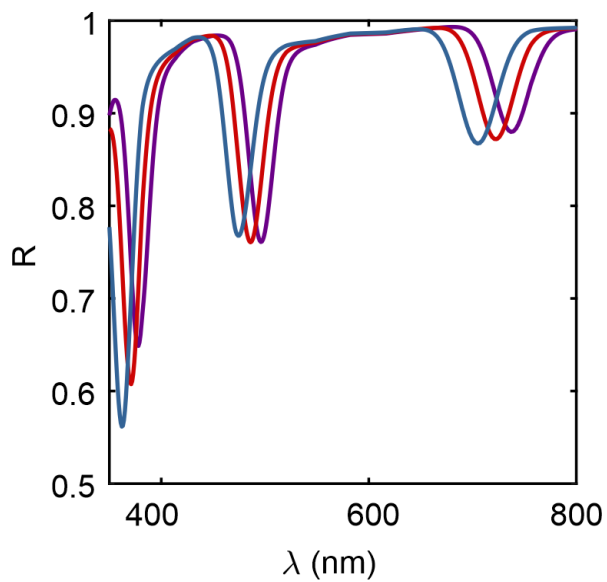

**Figure S2.** Reflectance spectra of the cavity at 26°, 36° and 46°(purple, red and blue, respectively) calculated using a constant index of 1.84.

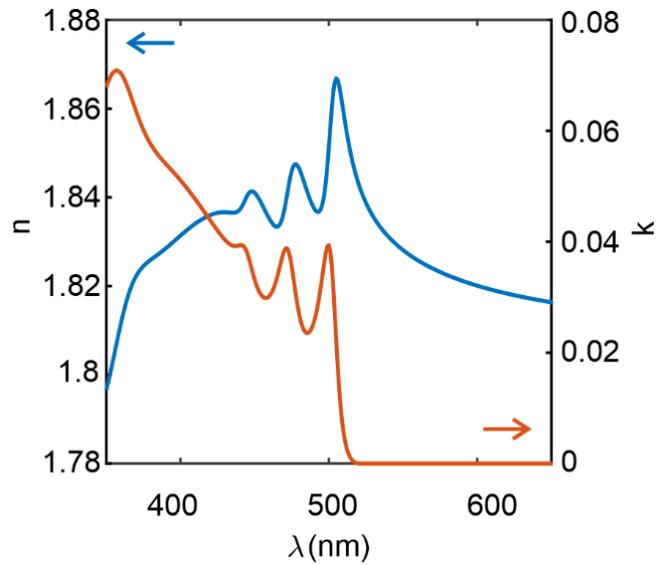

**Figure S3.** Real (n) and imaginary (k) parts of the refractive index of a film of CsPbBr$_3$ QDs with PS in blue and orange respectively.

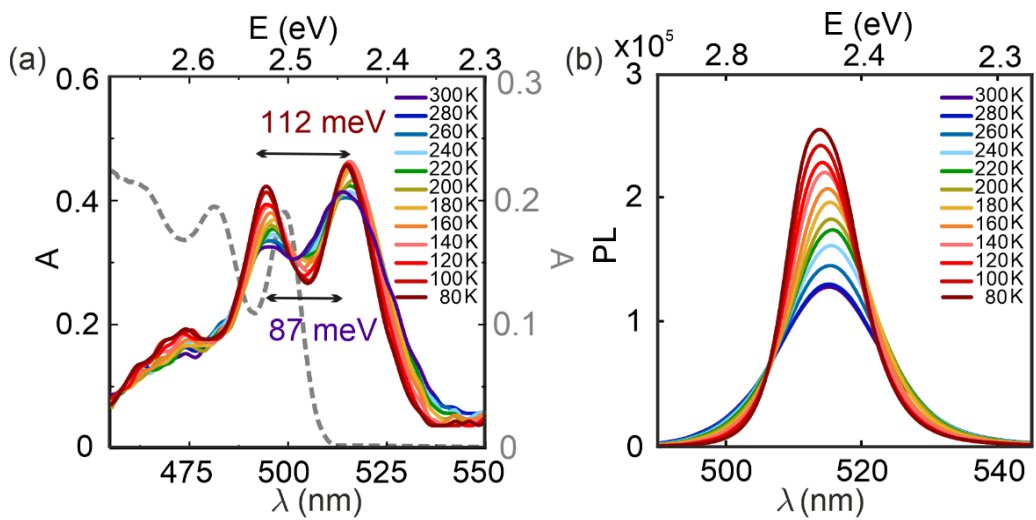

**Figure S4.** Absorptance (a) and photoluminescence (b) spectra of the CsPbBr$_3$ QDs + PS cavity measured from 300K to 80K.

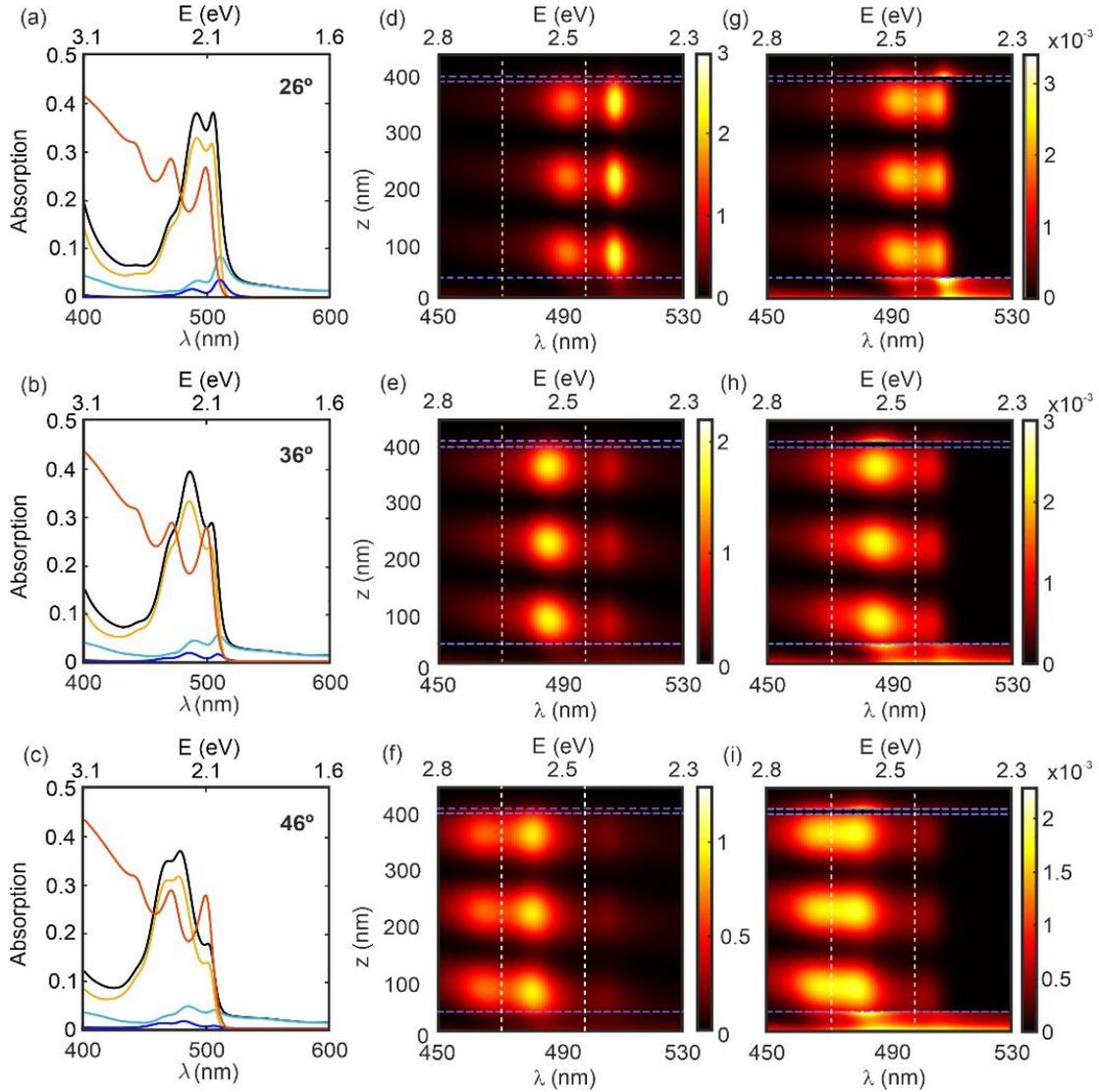

**Figure S5**. Visualization of optical mode splitting in CsPbBr$_3$-QDs solids. (a), (b) and (c), transfer matrix calculated spectra of the absorptance occurring at each layer in the ensemble, namely, top Ag (light blue line), CsPbBr$_3$·QDs (yellow line), bottom Ag (dark blue line) for 26°, 36° and 46° incidence angles, respectively. Black and orange lines are the absorptance of the whole cavity and a bare CsPbBr$_3$-QDs film with the same thickness at the same angles of incidence. (d), (e) and (f), calculated spatial and spectral profiles of the electric field intensity $|\mathbf{E(r)}|^2$, and, (g), (h) and (i), the absorbed luminous power P$_A$ for the CsPbBr$_3$ cavity, for incidence angles of 26°, 36° and 46°. Calculations are performed considering a plane wave impinging on the top silver mirror (position 0 in the z-axis) and propagating along the z-direction. Interfaces between layers are indicated by horizontal dashed blue lines, while vertical white dashed lines represent the position of the 1s-1s and 1p-1p excitons. In this representation, both the order of the optical mode participating in the coupling and its splitting are evident by the number of nodes observed in the spatial profiles of both $|\mathbf{E(r)}|^2$ and P$_A$.

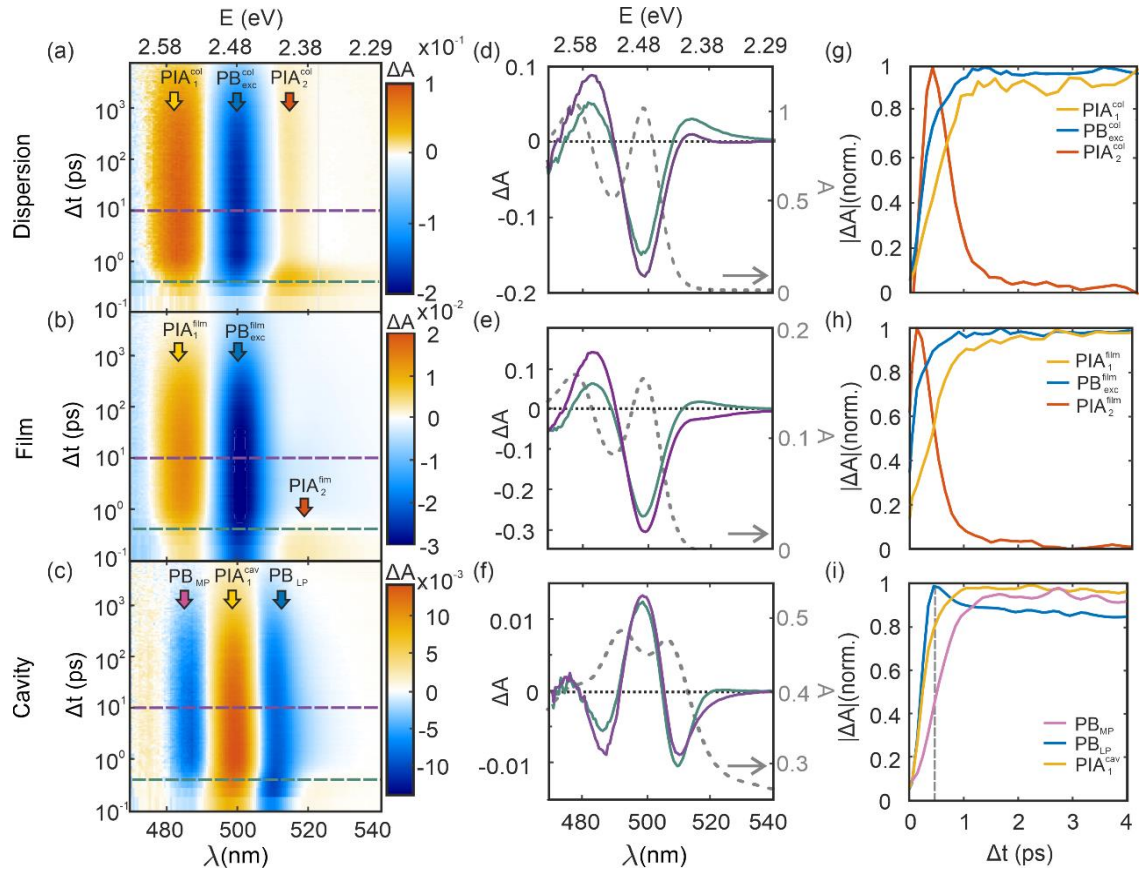

**Figure S6**. **Ultrafast transient absorption spectroscopy of CsPbBr₃-QDs dispersion, film and optical cavity**. Maps of ΔA versus $\lambda$ and $\Delta t$ for the CsPbBr₃-QD (a) dispersion, (b) bare film and (c) cavity, in which the main signals are labelled. Some illustrative selected ΔA spectra, attained at $\Delta t$=0.4 ps (green lines) and $\Delta t$=10ps (purple lines), are shown in (d), (e) and (f), respectively. Both $\Delta t$ are highlighted by dashed horizontal lines in panels (a), (b) and (c). The corresponding linear absorptance spectra are also plotted (grey dashed lines). Panels (g), (h) and (i) display the early-time evolution of the maximum intensity of the main signals extracted from the analysis of |ΔA|. Pump wavelength and duration used for these experiments were $\lambda$=420 nm and 190 fs; fluences were varied in the range 12-118 μJ/cm², depending on the sample absorptance at $\lambda$=420 nm, to ensure we have less than one excitation per PQD (please see section 3 in these Supplementary Methods,). Vertical grey dashed line in (i) indicates de maximum intensity of the LP photobleaching signal, which occurs at 0.45 ps. All cavity measurements were performed with an excitation and collection an angle of 26˚.

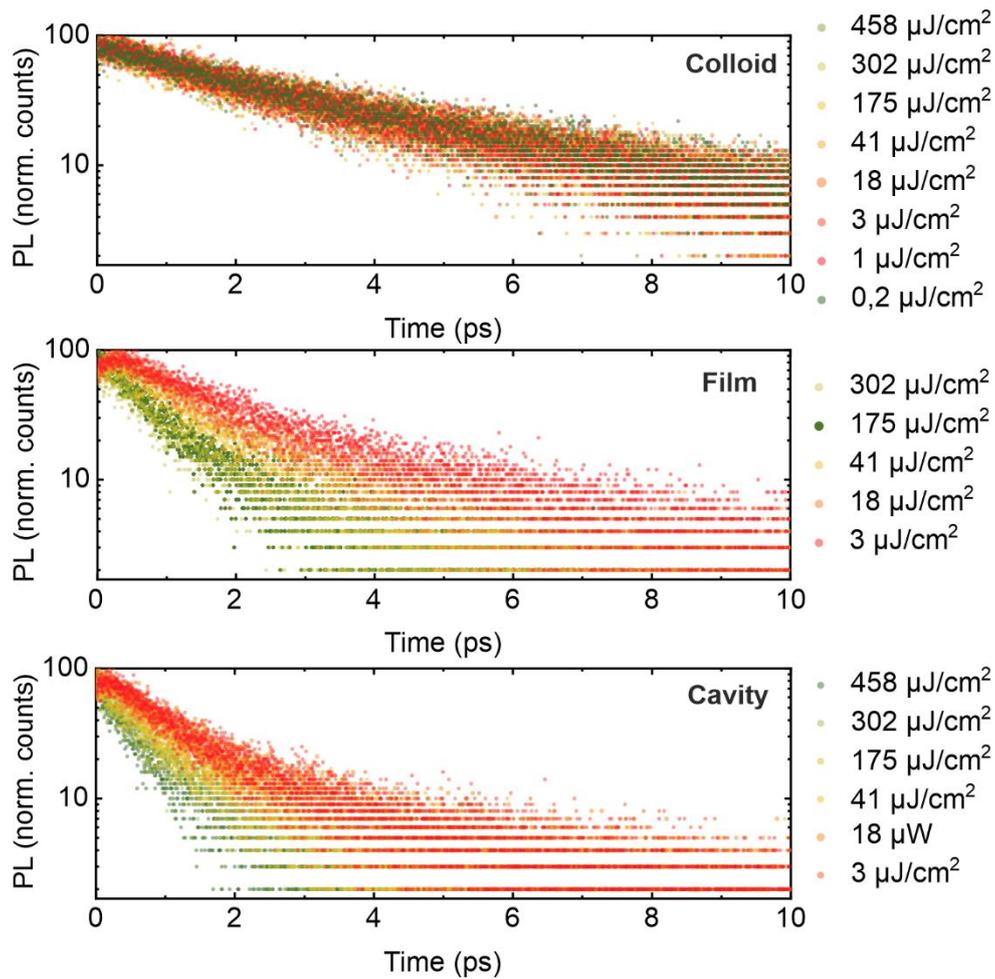

**Figure S7.** Photoluminescence vs. time of the three systems (colloid, film and cavity) at different fluencies with λ=420nm.

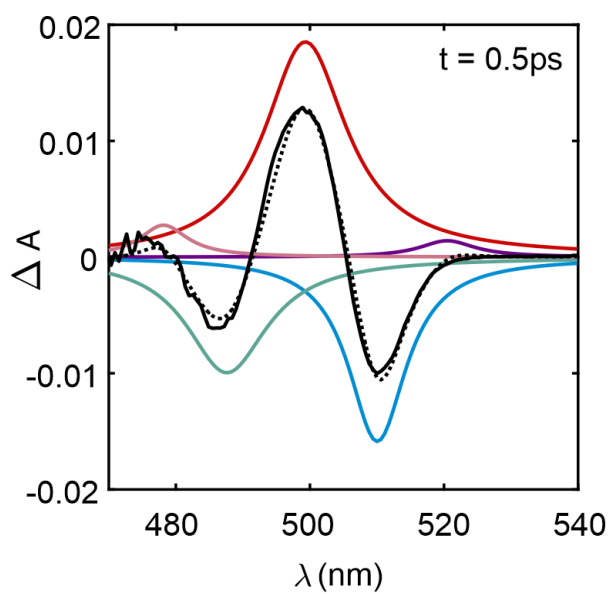

**Figure S8.** Fitting of five Voigt profiles to the experimental ΔA spectrum at 0.5ps.

**Supplementary Methods**

## 1. Refractive index and reflectance of the underlying cavity

The effective refractive index used to calculate the underlying cavity modes, i.e. those that would be observed if there was a non-absorbing layer between the mirrors, is estimated from the experimentally estimated n and k curves shown in Figure S2. From this, we set 1.84 as constant refractive index for the non-absorbing cavity, as it is the average value attained at the spectral range at which the polaritonic modes are observed. Also, in all cavity reflectance/absorptance calculations, the variations of the PQD layer thickness observed by electron microscopy are accounted for by averaging a gaussian distribution centered at 356 nm with a 25 nm FWHM.

## 2. Tavis-Cummings Hamiltonian solutions: polaritons, dark states and Hopfield coefficients

Within the single excitation subspace, strong coupling between *N* electronic transitions and one photon resonant mode leads to the formation of *N+1* coherent superpositions of excitations. Since the cavity dominates the response under external driving, and having the lower and upper hybrid light-matter states the highest contribution of the cavity mode, the absorption spectrum of the strongly coupled system shows mainly two peaks, corresponding to the lower and upper polaritons. The other *N*–1 levels (i.e., the vast majority of solutions), whose energies lie in between those of the upper and lower polaritons, cannot be accessed from the ground state for symmetry reasons and are referred to as dark states. In our case, we have *N* CsPbBr$_3$-QDs and two excitonic transitions per PQD, $\hbar\omega_s$ and $\hbar\omega_p$, involved in the coupling, which yields 2*N*+1 coherent superpositions of excitations, of which 2*N*-2 are **dark states** and 3 are the lower, middle and upper polaritons. With this in mind, we can obtain relevant information on the superposition of states leading to the formation of these polaritons by solving a simplified eigenvalue equation for the Tavis-Cummings Hamiltonian, $\hat{H}_{TC}$:

$$\hat{H}_{TC}\psi_{L,M,U} = \hbar\omega_{L,M,U}\,\psi_{L,M,U} \quad \text{(S1)}$$

The new hybrid states, $\psi_L$, $\psi_M$, and $\psi_U$, are the wavefunctions of the LP, MP and UP states, which result from the hybridization of the cavity photon and the two excitonic transitions involved in the coupling, described by the wavefunctions $\psi_\gamma, \psi_s, \psi_p$:

$$\psi_{L,M,U} = C_\gamma^{L,M,U}\psi_\gamma + C_s^{L,M,U}\psi_s + C_p^{L,M,U}\psi_p \text{(S2)}$$

In equation (S2), $C_i^j$ are the **Hopfield coefficients**, which fulfill the normalization condition:

$$\left|C_\gamma^j\right|^2 + \left|C_s^j\right|^2 + \left|C_p^j\right|^2 = 1 \quad\quad \text{(S3)}$$

On this grounds, Eq.(S1) becomes:

$$\begin{pmatrix} \hbar\omega_\gamma & \hbar\dfrac{\Omega_{\gamma,s}}{2} & \hbar\dfrac{\Omega_{\gamma,p}}{2} \\ \hbar\dfrac{\Omega_{\gamma,s}}{2} & \hbar\omega_s & 0 \\ \hbar\dfrac{\Omega_{\gamma,p}}{2} & 0 & \hbar\omega_p \end{pmatrix} \begin{pmatrix} C_\gamma \\ C_s \\ c_p \end{pmatrix} = \hbar\omega_{L,M,U} \begin{pmatrix} C_\gamma \\ C_s \\ c_p \end{pmatrix} \quad\quad \text{(S4)}$$

By solving this three coupled oscillator Hamiltonian, we attain the polariton energy dispersion curves, $\hbar\omega_{L,M,U}$ vs. $\mathbf{k}_{\parallel}$, which are plotted as white solid lines in Fig. 2a-2b in the main body of the manuscript, and the the Hopfield coefficients $\left|C_i^j\right|^2$, which give us the degree of contribution of each state to the different exciton-polariton states observed, as shown also in Figs. 2c-2e. Please note that in Eq.(S4), the number of identical excitonic transitions, N, coupled to the optical mode is included inside the coupling (off-diagonal) terms, through the Rabi frequencies $\Omega_{\gamma,s} = \sqrt{N_s}g_{\gamma,s}$, and $\Omega_{\gamma,p} = \sqrt{N_p}g_{\gamma,p}$, where $g$ is the coupling strength of that particular transition. $\omega_{exc}$ and $\omega_{mode}$ represent the exciton and photonic mode original frequencies

It should be noted that the description in terms of N particles and 1 mode is only a fair approximation to describe a planar cavity with a continuum of in-plane wave vectors k and the associated dispersion (i.e., the angle-dependent frequency). A more rigorous discussion of these systems may be made in terms of the density of states of molecular vs cavity excitations, as, e.g., in ref.[1].

### 3. Charge carrier density estimation in TAS measurements

The average value of the pump power (P) is measured at the sample position and it is used to calculate the fluency (fl) of a single pulse:

$$fl = \frac{P}{f \cdot \pi r^2} \qquad (5)$$

Where $f = 1kHz$ is the frequency of the pump pulses and $r = 100 \ \mu m$ is the radius of the spot illuminating the sample.

The number of charge carriers per unit volume ($n_{exc}$) generated after each pump pulse per NCs is calculated using the formula described by Savill et. al.[2] that considers a mean density inside the sample calculating the average between the charge carriers that would be generated at the top surface of the sample and the ones at the end of the sample after the absorption of a fraction of the incoming photons:

$$n_{exc} = \frac{fl(1-e^{-\alpha \cdot d \cdot ff})}{d \cdot ff \cdot E_{photon}} \qquad (6)$$

being $\alpha$ the absorption coefficient of the QDs, $d$ the thickness of the sample, $ff$ the filling fraction or the fraction of QDs volume in the sample (assuming the whole volume is the sum of ligands, polystyrene and QDs) and $E_{photon}$ the energy of one photon with the enery of the pump (420nm).This quantity multiplied by the volume of a nanocrystal, which it is assumed to be a sphere with a 6.6nm radius, gives us the charge carrier density per PQDl, $N_{exc}$.

### 4. Transient absorption data treatment

To analyse the time evolution of the cavity absorptance at resonance with 1s-1s exciton (at 26°) both the spectra of ΔA measured with ultrafast absorption techniques and the linear absorption modified by ΔA were separated in their main components. ΔA stands for the logarithmic ratio of change in absorption after the pump pulse. As the cavity transmittance is negligible in the spectrum range examined, the ΔA is measured by only considering the reflectance:

$$\Delta A = -\log(\frac{R_0}{R_{exc}}) \qquad (7)$$

Where $R_0$ is the linear reflectance and $R_{exc}$ is the reflectance measured after the pump (note that for the transmitting samples i.e. the colloidal dispersion and the film the reflectance is negligible and the signal is detected in transmission, redefining ΔA as $\Delta A = -\log(\frac{T_0}{T_{exc}})$ ), which is dependent on the time delay with respect to it. The absorptance spectrum after excitation is then calculated from the linear reflectance and the transient absorption spectrum as:

$$A_{exc} = 1 - R_{exc} = 1 - R_0 \cdot 10^{-\Delta A} \qquad (8)$$

To do so, the ΔA spectra was deconvoluted at each time into five contributions by the fitting of five Voigt functions whose parameters were obtained with the Matlab fmincon optimization function. The five functions (depicted in Fig. S6) correspond to a short lived (<1ps) photoinduced absorption (PIA) at around 520nm, two photobleaching (PB) at approximately the positions of the polaritons, a PIA between the polaritons positions and a low intense PIA at high energy, where the limit of the detector introduces some noise in the measured spectra. The linear absorptance spectrum modified at each time by ΔA was deconvoluted in three Voigt oscillators (Fig. 3i, in the main body of the manuscript) corresponding to the LP, MP and UP, although the last one is fixed in time due to the small interaction with the photonic mode at the angle of measurement, which results in a negligible variation of the absorptance with time.

## 5. Comparative analysis of transient absorption spectroscopy results for PQD colloidal dispersions, thin films and optical cavities.

Polariton excitation and decay dynamics in CsPbBr₃-QDs optical microcavities were studied by ultrafast transient absorption spectroscopy (TAS) using a rare-earth based femtosecond pulsed laser (pulse duration, 190 fs), whose beams were redirected through either an optical parameter amplifier, to generate the pump pulses, or a delay line and a sapphire crystal, to create broadband probe pulses. For the sake of comparison, a similar analysis was performed for the CsPbBr₃-QDs colloidal dispersion and the bare film. Considering the different absorptance of the three systems under analysis, the excitation photon fluence was varied in the range 12-118 μJ/cm² to achieve less than one excitation per QD, preventing the damage of the samples. Results are shown in Fig. S6, in which the full series of ΔA spectra (with ΔA being the result of subtracting the linear absorptance from that of the photoexcited sample) attained over a five order of magnitude time scale are plotted as intensity maps as a function of probe photon wavelength and pump-probe delay, *Δt*, following the non-resonant excitation (i.e., matching nor the excitonic transitions observed in CsPbBr₃-QDs colloidal dispersion and film, neither the polaritonic transitions in the cavity) at time 0 by a *λ*=420 nm pump pulse. Results attained for the PQD colloid, film and optical cavity are plotted in Figs. S6a, S6b and S6c, respectively. Selected ΔA spectra attained at different delay times are explicitly shown in Figs. S6d-S6f, namely at *Δt*=0.4 ps (green line) and *Δt*=10 ps (purple line). The early-time dynamics of the main signals identified in Figs. S6a-S6c are plotted in Figs. S6g-S6i, respectively.

The ultrafast response of both the CsPbBr₃-QDs colloidal dispersion and film show a prominent photobleaching of the first excitonic transition (PB$_{exc}^{col}$ and PB$_{exc}^{film}$, centred at 2.48 eV, shaded in blue) and a conspicuous photoinduced absorption (PIA$_1^{col}$ and PIA$_1^{film}$centred at 2.56 eV, shaded in yellow) spectrally located in between the first and second excitonic transitions, as the comparison between linear absorptance (dashed line) and TAS curves (green and purple solid lines) in Figs. S6d and S6e explicitly reveals. A third less intense and short-lived (<2 ps) signal

(PIA$_2^{col}$ and PIA$_2^{film}$, located around 2.40 eV, shaded in orange) is observed at energies right below the first excitonic transition. The PB$_{exc}$ signal is typical of semiconductor nanocrystals and is understood as the result of the filling of the lowest energy excited states as a consequence of the cooling of photocarriers.[3] The short-lived PIA$_2$ signal, also characteristic of QDs, is usually attributed to a slight red-shift of the 1s-1s transition caused by the Coulomb interaction between photoexcited e$^-$-h$^+$ pairs, the so-called biexciton effect.[4-6] As cooling takes place, hot carriers decay and fill in these newly available low energy states, thus extinguishing the PIA$_2$ signal.[7,8] The origin of the intense PIA$_1$ signal observed in CsPbBr$_3$-QDs, however, has been the subject of debate.[9,] It has been described as a distinctive feature of lead halide PQDs, and attributed to a parity-forbidden transition,[11] only observable due to the relaxation of the selection rules resulting from the formation of large polarons, i.e., significant exciton-induced deformations of the [PbBr$_3$]$^-$ sublattice.[12,13] In this context, it has been proposed that large polarons in CsPbBr$_3$-QDs could be also contributing to the screening of the Coulomb interaction between e$^-$-h$^+$ pairs,[14] as the coincident dynamics of the PIA$_2$ decay (orange lines in Figs. S6g and S6h) and of the PIA$_1$ rise (yellow lines) seem to indicate. On the other hand, recent studies alternatively assign the PIA$_1$ signal to higher biexcitonic transitions,[15] a hypothesis supported by detailed calculations of the spectrum of confined exciton-to-biexciton transitions in CsPbBr$_3$ lattices. In either case, the PIA$_1$ signal is considered to be the result of transitions that become available only upon photoexcitation.

Bearing in mind the response of the bare CsPbBr$_3$-QD film, the ultrafast $\Delta A$ signal from the optical cavity displayed in Fig. S6c was analysed. Two intense bleaching signals are detected at the lower (PB$_{LP}$) and middle (PB$_{MP}$) polariton spectral positions (located at 2.43 eV and 2.55 eV, respectively). Also, a very intense photoinduced absorption (PIA$_1^{cav}$, at 2.49 eV) is observed between PB$_{LP}$ and PB$_{MP}$. The early stage dynamics of these signals are compared in Fig. S6i. In there, it can be seen that PB$_{LP}$ and PB$_{MP}$ signals (blue and pink line respectively) present a very different rise time ($\tau_{PB,LP}$<0.5 ps, $\tau_{PB,MP}$<2 ps). Although the bleaching of the lower polariton is the consequence of the characteristic filling of the lowest energy states due to hot-carrier relaxation, the subsequent long-lived bleaching of the middle polariton cannot be understood without considering the interplay with the abovementioned large reservoir of dark states. Even though not directly accessible from the ground state, dark states can be filled both from the lower polariton state and from higher energy levels as hot electrons cool down.[16] This picture is further supported by the partial recovery of the absorption evidenced by the peak observed in the PB$_{LP}$ signal at $\Delta t$=0.45 ps (signposted by a vertical grey dashed line in Fig. S6i), which points at a transfer of carriers from the lower polariton state to the dark state reservoir once a certain occupation level is attained. Thus, gradual filling of the dark states could eventually lead to transfer of carriers to the middle polariton state, giving rise to its bleaching. These carrier exchanges are favoured by the significant overlap between dark and polaritonic states, which is expected in our case due to the moderate separation observed between lower and middle polaritons.[17,18] Interestingly, the low energy photoinduced absorption observed in the dispersion and the film, PIA$_2$, attributed to the shift of the lowest excited state energy caused by biexciton interaction, is almost imperceptible in the cavity, which indicates that no significant renormalization is taking place in the cavity.

## 6. Log-normal fit of PL decay curves

PL decay curves were fitted using an exponential model weighted by with a gaussian distribution of decay rates (log-normal function)[19]. PL counts at each time are then given by

$$\int_0^\infty \rho(\Gamma) \exp(-\Gamma t) d\Gamma \qquad (9)$$

where $\Gamma$ is the decay rate and $\rho(\Gamma)$ is the gaussian distribution of decay rates:

$$\rho(\Gamma) = C \frac{1}{\sigma \Gamma \sqrt{2\pi}} \exp\left[-\frac{1}{2}\left(\frac{ln\Gamma - \mu}{\sigma}\right)^2\right] \qquad (10)$$

Here, $C$ is an amplitude factor, $\sigma$ is the standard deviation and $\mu$ the mean of the logarithm of $\Gamma$. Fitting to experimental data was performed using Matlab non-linear regression function *Nlinfit* to estimate the parameters $C$, $\mu$ and $\sigma$ with the model described in expressions (9) and (10).